\documentclass[aps,pre,twocolumn,showkeys,amssymb,floatfix]{revtex4}
\usepackage{graphicx}
\usepackage{epsf}

\begin{document}

\title{Minorities in a model for opinion formation}

\author{M. F. Laguna, Guillermo Abramson and Dami{\'a}n H. Zanette}

\affiliation{Centro At{\'o}mico Bariloche, CONICET and Instituto
Balseiro, 8400 San Carlos de Bariloche, R{\'\i}o Negro, Argentina}

\date{\today}

\begin{abstract}
We study a model for social influence in which the agents' opinion
is a continuous variable [G. Weisbuch et al., Complexity
\textbf{7}, 2, 55 (2002)]. The convergent opinion adjustment
process takes place as a result of random binary encounters
whenever the difference between agents' opinions is below a given
threshold. The inhomogeneity in the dynamics gives rise to a
complex steady state structure, which is also highly dependent on
the threshold and on the convergence parameter of the model.
\end{abstract}


\maketitle

\section{INTRODUCTION}

The formation of public opinion is a complex phenomenon that
affects almost all aspects of human life. In a population where
each agent has an independent opinion, a typical situation when a
common decision needs to be taken involves the exchange of
information and points of view. As a result of these exchanges,
people gradually change their opinions and a collective state
emerges. In some cases, a single position is held by the
population in the end. In others, a state of coexistence between
diverging opinions may result and persist. In some regards, the
emergence of a collective state is reminiscent of many other
synchronization effects in populations with global coupling in
the fields of physics, chemistry and biology
\cite{mikhailov,kuramoto,winfree}. The starting point of the
present work is a model for   opinion formation introduced by
Weisbuch and coworkers \cite{weisbuch}.
The state of the agents is a continuous variable, representing an
opinion, that evolves according to rules tending to homogenize
the system. The conditions to reach consensus in this model were
analyzed in previous works \cite{weisbuch,stone}, but several
interesting properties were not studied. In particular, we analyze
the transient state during which the formation of the collective
opinion takes place, as well as the relevance of dynamical time scales
in the formation of minorities. As an alternative to continuous
opinions, discrete opinions have also been studied by Weisbuch and
coworkers \cite{weisbuch}, and by ourselves in a previous work
\cite{laguna}.

Let us consider a population of $N$ agents, each of them
characterized by a continuous state variable, the ``opinion,''
$x\in [0,1]$. Agents interact and update their opinions as a
result of mutual influence in the following manner. At each time
step $t$, two randomly chosen agents meet. If the difference
between their opinions is smaller than a threshold,
$|x(t)-x'(t)|<u$, both change their opinions according to the
rule:
\begin{eqnarray}
x(t+1)&=&x(t)+\mu  [x'(t)- x(t)], \label{dynamics1}\\
x'(t+1)&=&x'(t)+\mu  [x(t) -x'(t)],
\label{dynamics2}
\end{eqnarray}
where $x$ and $x'$ are the agents' opinions and $0<\mu<1$ is a
convergence parameter \cite{weisbuch}. That is, if the two
opinions are already sufficiently close, they approach even more.
On the contrary, if the two opinions are not close enough, the
agents keep their opinions unchanged. The convergence parameter
$\mu$ controls the approach of the two differing opinions. The
threshold $u$ is taken as constant in time and uniform across the
whole population. Weisbuch and coworkers have shown that, for
large threshold values ($u>0.3$), consensus is achieved in the
system, whereas for lower values of $u$ several opinions persist,
the number of which follows a general functional dependence of
the form $[1/2u]$ on the threshold, where $[x]$ indicates the
integer part of $x$. The role of $\mu$, they observe, is only that
of controlling the speed of convergence towards the stationary
state. Below, we explore a further role of this parameter.

The final product of the process is a set of clusters of opinion,
each one of them consisting of a subpopulation of agents with
identical opinions. This clusterization process is gradual and in
principle achieved at infinitely long times, due to the continuous
character of the opinion variable and the exponential approach
consequence of Eqs. (\ref{dynamics1}) and  (\ref{dynamics2}).
Nevertheless, the formation of the final clusters is evident at
finite times, because the distance between clusters is bounded
from below by the threshold $u$. It is then possible to define a
criterion for the formation of clusters as follows: two agents
belong to the same cluster of opinions if the difference between
them is less than certain finite value $u_0$, smaller than $u$. A
criterion like this is also sensible from a realistic point of
view, since it would be unreasonable to maintain that two
opinions could differ in an arbitrarily small quantity. In the
results shown in the present paper, we fix $u_0=0.001$. This value
of $u_0$ limits the number of effective opinions to one thousand,
which is large enough for the systems with $N=10^3$ to $5\times
10^5$ used in our simulations.

The formation of the clusters, with the criterion given above, is
complete at a time $t_f$ proportional to $N$. The simulation is
stopped when no more activity is observed in the clusterization
process. The final state consists of a number of ``major''
clusters and a number of ``minorities,'' defined as clusters with
less than 10\% of the population. As will be discussed below, the
major clusters are determined by the value of the threshold. These
are the clusters analyzed by Weisbuch and coworkers \cite{weisbuch}.
The minorities---which, we stress, are not an artifact of the
discretization criterion imposed on the opinion space---are a
byproduct  of the dynamics, determined by the parameter $\mu$,
through the dependence of the final state on the initial condition
and on the transient steps.

\section{NUMERICAL RESULTS}

\begin{figure}[tbp]
\centering
\resizebox{\columnwidth}{!}{\includegraphics[40,260][720,800]{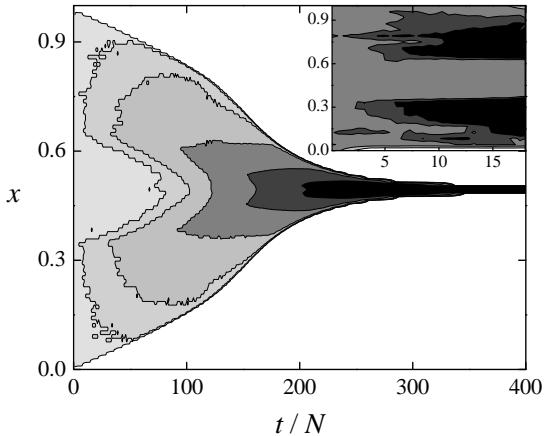}}
\caption{Distribution of opinions as a function of time. The
number of agents at each point of the graph is indicated in gray
scale, with darker tones corresponding to higher densities. The inset
shows in detail the first stage of the evolution, where the
merging of the initial clusters can be seen. $N=5\times 10^5$,
$u=0.35$, $\mu=0.001$. The scale of grays is different in both
plots, to facilitate the reading.}
\label{fig1}
\end{figure}

We analyze the dynamical behavior of the system starting from a
random  uniform distribution of opinions in the interval $[0,1]$.
We begin with a
value of the threshold large enough to induce the formation of a
single major cluster. As time progresses, the region in opinion
space occupied by the system shrinks, as shown in the time charts
of Ref.~\cite{weisbuch}. Those plots, however, do not show the number
of agents with the same value of opinion (the population of the
clusters), which gives additional information about the
clusterization process. To illustrate this feature of the evolution
process, we plot in Fig. \ref{fig1} the distribution of opinions as
a function of time for a system with $N=5\times 10^5$, $u = 0.35$ and
$\mu=0.001$. We have chosen this very small value of the convergence
parameter to allow for access to the very initial steps of clusterization.
In the figure, a density plot shows with darker colors the regions
with more populated clusters.

In the inset of Fig. \ref{fig1} we show the
initial stage of the clusterization process. It can be seen that
clusterization proceeds very fast to the formation of two
main branches of opinion by the merging of smaller clusters,
that appear as horizontal streaks in the inset. After this, the
two main branches merge into the single final cluster, as shown
in the main plot of Fig. \ref{fig1}.

\begin{figure}[b]
\centering \resizebox{\columnwidth}{!}{\includegraphics{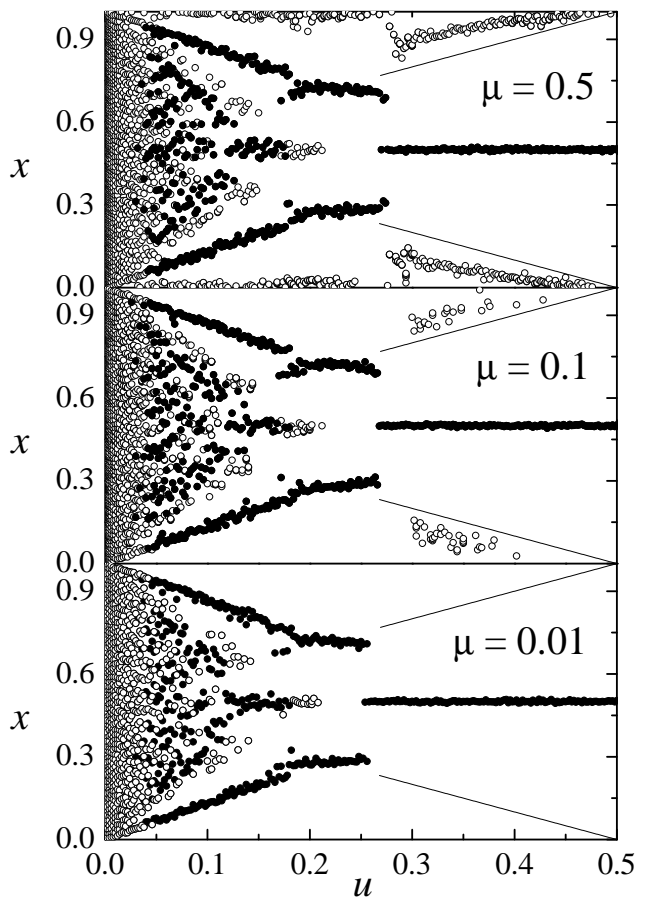}}
\caption{Distribution of agent's opinions as a function of the
threshold for three values of the convergence parameter, indicated
in the figures. Empty dots indicate minor clusters, whereas
full dots indicate major clusters, with population larger than
1000 agents, i.e. than 10\% of the total population. The lines
enclose the basin of attraction of the state with a single cluster.
The results of a single realization are shown in each panel for each value of $u$.}
\label{fig2}
\end{figure}

The final state can be characterized by the number of clusters of
opinion. For $u>0.3$ only one cluster is observed, whereas for
lower values of $u$ several clusters persist, as studied in Ref.
\cite{weisbuch}. This situation takes place despite  the fact
that an opinion can propagate through the whole population. We have
found that the final state depends, further, on the convergence speed
through the parameter $\mu$. In Fig. \ref{fig2} we plot the distribution
of opinions as a function of the threshold for three typical values
of $\mu$. We distinguish major clusters, with a population larger
than $10\%$ of the total (full dots), from minor clusters (empty
dots). While the structure of the major clusters is the same in
all the cases, the density of minor clusters near the extreme
opinions $0$ and $1$ is higher for the larger values of $\mu$.

\begin{figure}[tbp]
\centering
\resizebox{\columnwidth}{!}{\includegraphics{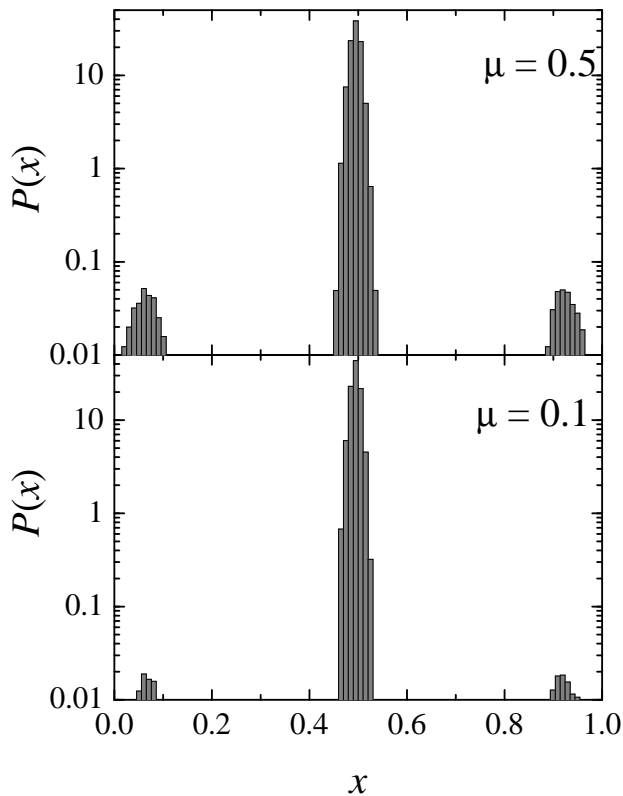}}
\caption{Histogram of opinion for $2000$ realizations with
$N=1000$ and a fixed threshold, $u=0.35$. The central peak in
both figures corresponds to major clusters, whereas the two
lateral peaks are formed by minor clusters. In the upper panel
the convergence parameter is big ($\mu=0.5$) whereas in the lower
panel $\mu=0.1$. Note the difference in the area of the minor
clusters. Also observe that the distance between the clusters may
be smaller than $u$, since the histograms are built from $2000$
independent realizations.}
\label{fig3}
\end{figure}

Numerical simulations performed over all the range of thresholds
show that the origin of the minor clusters is strongly dependent
on the speed of the dynamics through the parameter $\mu$. For a
given value of $u$, agents lying within a ``basin of attraction''
of radius $u$ from the budding clusters at an early stage of the
evolution are almost certainly attracted to them, with a speed
depending on $\mu$. In Fig. \ref{fig2} we show the basin of
attraction of the central opinion in the range of $u$ where a
single cluster is obtained; the basins corresponding to states
with more clusters are omitted to avoid clutter. Agents lying
outside this basins cannot interact directly with the agents
forming the clusters---since their opinions differ by more than
$u$. However, agents with intermediate opinions may serve as
bridges during the transient evolution, allowing agents from
outside the basins to enter them. The availability of these
intermediate agents strongly depends on $\mu$, since they
ultimately tend to approach the clusters at a speed given by
$\mu$. As a consequence the number of agents that  remain frozen
outside the basins increases with $\mu$. These agents,
nevertheless, can interact with each other, and finally form a
cluster by themselves, albeit a minor cluster, not a major one.
These are the agents that appear outside the basins, in Fig.
\ref{fig2}, for $\mu=0.1$ and $0.5$. The bottom panel of Fig.
\ref{fig2}, corresponding to the slowest dynamics at $\mu=0.01$,
does not exhibit these extreme opinions: the transient evolution
has been slow enough to allow all of them to be incorporated into
the central clusters.

The population of the minor clusters can be better appreciated in
Fig. \ref{fig3}, where the results of 2000 independent realizations
are averaged, for a fixed value of $u=0.35$ (one major cluster). When
the convergence parameter is large (Fig. \ref{fig3}, top), the
lateral clusters are, indeed, much bigger than the corresponding
ones in the bottom plot, where the dynamics is slow and only a
few agents remain outside the major cluster at the central
opinion. It is evident, then, that the role of the parameter
$\mu$ goes beyond than that of a speed of convergence parameter.
It is able to modify the equilibrium state of the system.

The minorities located near the central opinion behave
differently from those that, like the ones shown in Fig.
\ref{fig3}, exist near the extremes. We can see in Fig.
\ref{fig2} that around $u=0.2$ there is a number of minorities
that persist even for very low values of $\mu$ (refer to the
empty dots near $x=0.5$ at $u\approx 0.2$). These clusters lie
outside the basins of the two clusters that attract most of the
population, so they are truly stationary. The intermediate
opinions fail to move them to one of the major clusters during
the early stages of the evolution because their central position
makes them equally probable to move to either side in each
individual interaction. Later, when the intermediate agents have
disappeared, these minorities persist, and this happens for any
value of the convergence parameter. The net effect of this in the
phase portrait is that the transition from three clusters to two
clusters results shifted towards larger values of $u$. The
phenomenological rule, found in Ref. \cite{weisbuch}, that the
number of clusters depends on $u$ as $[1/2u]$, predicts a value
of $u=1/6$ for this transition, lower than $u\approx 0.21$, as
found in our results (Fig. \ref{fig2}). Certainly, this
phenomenon occurs at lower thresholds also, affecting the other
transitions as well and in the same direction. Its observation
is, however, very difficult when the number of clusters is larger
than three.

\begin{figure}[tbp]
\centering
\resizebox{\columnwidth}{!}{\includegraphics[1,300][800,750]{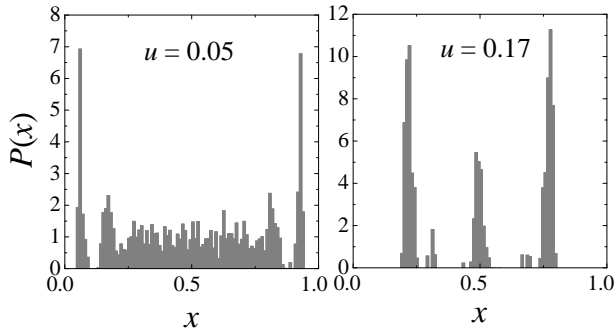}}
\caption{Histograms of opinion for 100 realizations and two
values of the threshold indicated in the figure, for $N=10^4$ and
$\mu=0.1$. Note the larger population of the exterior clusters. }
\label{fig4}
\end{figure}

At lower values of $u$, where the stationary phase consists of
more than two major clusters, it becomes manifest that the space
of opinions is heterogeneous, i.e. the dynamics of an agent with
a extremist opinion is different from the one with an
intermediate opinion. In Fig. \ref{fig4} we plot a histogram of
opinions for 100 realizations and two values of the threshold,
$u=0.17$ (three clusters), and $u=0.05$ (many clusters). In both
cases, extremist clusters are overpopulated.  Extreme clusters
can be populated by the extremist agents that do not have access
to the central clusters because of the small value of the threshold.
This gives advantage to the extreme clusters. This feature is not
present in other models of opinion space, such as binary models
\cite{laguna}, where there are no privileged opinions in this
regard.

\section{CONCLUSION}

We have analyzed in detail a model of social influence and public
opinion formation, previously presented by Weisbuch et al.
\cite{weisbuch}. The model treats opinion as a continuous scalar
quantity, a simplification that can be of relevance in the
process of decision making of very specific matters. The dynamics
of the agents is a mechanism that approaches already nearby
opinions, following Axelrod's \cite{axelrod} seminal ideas in the
field of dissemination of culture. Our simulations confirm
previous results for the model: large
thresholds produce consensus in the population, while low
thresholds do not, even when the randomness of the initial
condition and the interaction mechanism
allow, in principle, for the propagation of opinions throughout the
whole population. Our new results provide details on the transient
clusterization process, on the structure of the stationary phase
controlled by the threshold parameter, and on the role of the
convergence parameter on the final distribution of opinions. Even
though we have not shown it, these results have proved to be very
robust against noise of rather high amplitude.

The transient phase towards the clustered state allowed by the
threshold shows  progressive clusterization of the initial
random state, by means of a collapse of the clusters that form in
close regions of opinion. The speed of this process is certainly
governed by the parameter $\mu$ which, besides, determines how
many agents are left ``frozen'' outside the major clusters. If
the dynamics proceeds slowly (small $\mu$), many agents whose
opinions are initially far from what will become the consensus
(farther than $u$, indeed), can nevertheless, by interacting with
intermediate agents, be enabled to join the rest. If, on the
other hand, the dynamics is fast, a large number of agents are
left outside the consensus and end up forming extremist clusters
of opinion.

As a result of this process, the final state of the system
consists of a number of major clusters arranged, as a function of
$u$, in the hierarchical structure shown in Fig.~\ref{fig2}. A
varying number of minorities, depending on the value of $\mu$, may
also appear in the spaces between the major clusters in the final
state. All clusters of opinion, major and minor, result more
populated if they reside near one of the extremes of the opinion
space. This is a consequence of the heterogeneity of opinion space
in this kind of models, where the agents with extreme opinions are
effectively less able to interact with others. For a given threshold
$u$,  agents with opinions in the middle range  have in contrast
more potential partners and tend to spread in the final state.
Even if simplified, it is worthwhile to stress that this happens
in many situations of collective decision making in human
affairs, where extreme positions tend to remain irreconcilable
with the majority.

In summary, despite its simplicity, the continuous opinion model
with convergent dynamics displays a richness of features that can
shed light over real system of opinion formation.

\section{ACKNOWLEDGMENTS}

M.F.L. thanks the Solid State Group of Centro At{\'o}mico Bariloche
for the use of computational facilities.

\end{document}